\documentclass[preprint,prd,nofootinbib,tightenlines,amsmath,amssymb]{revtex4}
\usepackage{enumerate}
\usepackage{multirow}
\usepackage[utf8]{inputenc}
\usepackage{anysize}
\usepackage{textcomp}
\usepackage{epsfig}
\usepackage{graphics,color} 
\usepackage{verbatim}
\usepackage{afterpage}
\usepackage{amsmath}    
\usepackage{soul}
\usepackage{xcolor,cancel}
\DeclareUnicodeCharacter{00A0}{ }

\usepackage{blindtext}

\catcode`\@=11
\def\sla@#1#2#3#4#5{{%
 \setbox\z@\hbox{$\m@th#4#5$}%
 \setbox\tw@\hbox{$\m@th#4#1$}%
 \dimen4\wd\ifdim\wd\z@<\wd\tw@\tw@\else\z@\fi
 \dimen@\ht\tw@
 \advance\dimen@-\dp\tw@ \advance\dimen@-\ht\z@
 \advance\dimen@\dp\z@
 \divide\dimen@\tw@ \advance\dimen@-#3\ht\tw@
 \advance\dimen@-#3\dp\tw@ \dimen@ii#2\wd\z@
 \raise-\dimen@\hbox to\dimen4{%
 \hss\kern\dimen@ii\box\tw@\kern-\dimen@ii\hss}%
 \llap{\hbox to\dimen4{\hss\box\z@\hss}}}}

\def\cpto{\mathrel {\vcenter {\baselineskip 0pt \kern 0pt
    \hbox{$H_{r.f.}$} \kern 0pt \hbox{$\longrightarrow$} }}}
\def\slashed#1{%
 \expandafter\ifx\csname sla@\string#1\endcsname\relax
{\mathpalette{\sla@/00}{#1}}
\fi}
\def\declareslashed#1#2#3#4#5{%
 \expandafter\def\csname sla@\string#5\endcsname{%
#1{\mathpalette{\sla@{#2}{#3}{#4}}{#5}}}}
 \catcode`\@=12
\declareslashed{}{/}{.08}{0}{D}
 \declareslashed{}{/}{.1}{0}{A}
 \declareslashed{}{/}{0}{-.05}{k}
 \declareslashed{}{/}{.1}{0}{\partial}
 \declareslashed{}{\not}{-.6}{0}{f}

\def\lsim{\mathrel {\vcenter {\baselineskip 0pt \kern 0pt
    \hbox{$<$} \kern 0pt \hbox{$\sim$} }}}
\def\gsim{\mathrel {\vcenter {\baselineskip 0pt \kern 0pt
    \hbox{$>$} \kern 0pt \hbox{$\sim$} }}}

\newcommand{\bea}{\begin{eqnarray}}
\newcommand{\eea}{\end{eqnarray}}

\begin{document}

\baselineskip=15pt
\preprint{}

\title{CP violation in $h\to \tau\tau$ and  LFV $h\to \mu\tau$}

\author{Alper Hayreter$^1$\footnote{Electronic address: alper.hayreter@ozyegin.edu.tr}, Xiao-Gang He$^{2,3,4}$\footnote{Electronic address: hexg@phys.ntu.edu.tw}, German Valencia$^{5}$\footnote{Electronic address: German.Valencia@monash.edu }}
\affiliation{
$^{1}$Department of Natural and Mathematical Sciences, Ozyegin University, 34794 Istanbul Turkey.\\
$^{2}$Physics Division, National Center for Theoretical Sciences, Hsinchu, Taiwan 30013.\\
$^{3}$INPAC,Department of Physics and Astronomy, Shanghai Jiao Tong University, Shanghai.\\
$^{4}$Department of Physics, National Taiwan University, Taipei. \\
$^{5}$School of Physics and Astronomy, Monash University, 3800 Melbourne Australia.\footnote{On leave from Department of Physics, Iowa State University, Ames, IA 50011.}
}

\date{\today}

\vskip 1cm
\begin{abstract}

The CMS collaboration has reported a possible lepton flavour violating (LFV) signal $h\to\mu\tau$. Whereas this does not happen in the standard model (SM), we point out that new physics responsible for this type of decay would, in general, also produce charge-parity (CP) violation in $h\to \tau\tau$. We estimate the size of this effect in a model independent manner and find that a large asymmetry, of order 25\%, is allowed by current constraints.

\end{abstract}

\pacs{PACS numbers: }

\maketitle

Within the SM the tau-lepton coupling to the Higgs boson is uniquely determined by its mass. The Yukawa Lagrangian for leptons reads 
\begin{eqnarray}
{\cal L}_{Y}&=& y_{ij} \bar{\ell}_{Li}\ e_{Rj}  \phi +\ {\rm h.c.}
\label{lyuk}
\end{eqnarray}
Here $\ell_{Li}$ is the left handed SM lepton doublet, $e_{Rj}$ the right handed  lepton singlet, $\phi$ is the scalar Higgs doublet and $i,j=1,2,3$ are generation indices. The leptons acquire a mass when electroweak symmetry is broken and the Higgs field develops a vacuum expectation value (vev) $\langle\phi\rangle=v/\sqrt{2}$, $v\approx 246$~GeV. Eq. (\ref{lyuk}) then takes the form
\begin{eqnarray}
{\cal L}_{Y}&=&\left(1+\frac{h}{v}\right)\frac{y_{ij} v}{\sqrt{2}}\bar{e}_{Li}\ e_{Rj}  +\ {\rm h.c.} 
\label{lyuk2}
\end{eqnarray}
The Yukawa Lagrangian in the lepton mass eigenstate basis is obtained from Eq. (\ref{lyuk2}) with a bi-unitary transformation $S_e^\dagger (v y_{ij}/\sqrt{2}) T_e = m_i \delta_{ij}$. In this basis 
the $h\ell\ell^\prime$ couplings are given by $g_{h\ell_i\ell_j} = m_i \delta_{ij}/v$. They are thus proportional to the lepton masses, flavour diagonal and real and therefore CP conserving. 
 
Beyond the standard model (BSM), however, this no longer holds. In a model independent manner we can describe new physics with an effective Lagrangian that respects the symmetries of the SM. Identifying  the 125~GeV state observed at LHC as the SM Higgs, and assuming that there are no new particles below a few hundred GeV, the appropriate effective Lagrangian for BSM physics  is that of Buchmuller and Wyler \cite{Buchmuller:1985jz,Grzadkowski:2010es}. At leading order, with operators of dimension six, one already finds terms in the Lagrangian that modify Eq. (\ref{lyuk}), for example,
\begin{eqnarray}
{\cal L}_{6}&=&\frac{g_{ij}}{\Lambda^2}\  ( \phi^\dagger \phi) \bar{\ell}_{Li}\  e_{Rj}\ \phi +\ {\rm h.c.}
\label{dim6}
\end{eqnarray}
The matrix $g_{ij}$ is, in general, non-diagonal and complex. Expanding this Lagrangian after electroweak symmetry breaking in combination with  Eq. (\ref{lyuk}) we find, 
\begin{eqnarray}
{\cal L}_{Y(4+6)}&=&\left(1+\frac{h}{v}\right)\frac{y_{ij} v}{\sqrt{2}}\bar{e}_{Li}\ e_{Rj}   +\ {\rm h.c.} \nonumber \\
&+& \frac{v^2}{2\Lambda^2}\left(1+\frac{3h}{v}\right)\frac{g_{ij} v}{\sqrt{2}}\bar{e}_{Li}\ e_{Rj}   +\ {\rm h.c.} 
\end{eqnarray}
There is a bi-unitary transformation that diagonalizes the mass terms,  
\begin{equation}
S_e^\dagger \frac{v }{\sqrt{2}}\left(y_{ij}+\frac{v^2}{2\Lambda^2}g_{ij}\right) T_e = m_i \delta_{ij}
\label{massm}
\end{equation}
but it no longer diagonalizes the $h\ell\ell^\prime$ couplings. The matrix
\begin{equation}  
S_e^\dagger \frac{1}{\sqrt{2}}\left(y_{ij}+\frac{3v^2}{2\Lambda^2}g_{ij}\right) T_e =\frac{m_i}{v} \delta_{ij} +\frac{v^2}{\sqrt{2}\Lambda^2}S_e^\dagger g_{ij} T_e 
\label{fulyuk}
\end{equation}
remains an arbitrary complex matrix. This can be easily checked, for example using the Fritzsch ansatz  \cite{Fritzsch:1979zq,Cheng:1987rs}  for the $y_{ij}$ and treating the $g_{ij}$ as small corrections as suggested by the prefactor $v^2/\Lambda^2$. To relate the magnitudes of $h\to \tau\mu$ and $h\to \tau\tau$ one needs a specific model for the $g_{ij}$ matrix,  this effective Lagrangian tells us that both are expected with a common suppression factor relative to the SM terms.

\begin{figure}[h]
\includegraphics[scale=0.7]{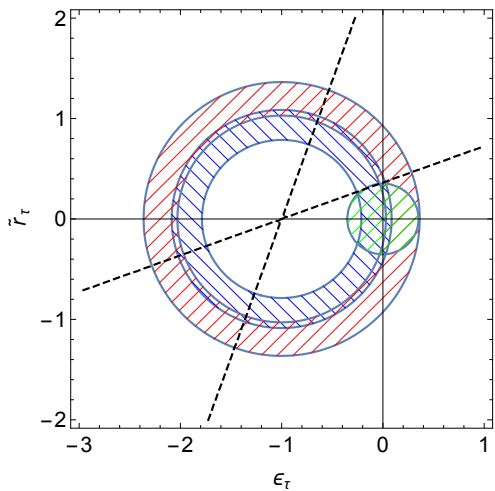} 
\caption{Region of parameter space allowed by the ATLAS (red) and CMS (blue) constraints in Eq. (\ref{expc}). The green dashed region illustrates the constraint from LFV Eq. (\ref{cmscon}). The diagonal black lines show the maximum value of $ (r_\tau \tilde{r}_\tau ) /(r_\tau^2 + \tilde{r}_\tau^2  )\approx 0.32$ allowed by these constraints.}
\label{fig}
\end{figure}

It has become customary to parametrize a generic tau-lepton Yukawa coupling as 
\begin{eqnarray}
g_{h\tau\tau}&=&-\frac{m_\tau}{v}h\bar\tau\left(r_\tau +i\tilde{r}_\tau \gamma_5\right)\tau  \label{thisL} 
\end{eqnarray}
and in terms of the deviations from the SM to further write $r_{\tau} = 1+\epsilon_{\tau} $. In terms of Eq. (\ref{fulyuk}), and simplifying our notation by writing
\begin{eqnarray}
 \frac{v^2}{\sqrt{2}\Lambda^2}(S_e^\dagger g_{ij} T_e)_{33} \equiv g_{\tau\tau}e^{i\alpha},
\end{eqnarray}
 then
\begin{eqnarray}
\epsilon_{\tau} =  \cos\alpha \ g_{\tau\tau}\ \frac{v}{m_\tau}, &&
\tilde{r}_{\tau} =  \sin\alpha  \ g_{\tau\tau}\ \frac{v}{m_\tau}.
\label{gtt}
\end{eqnarray}
The lepton-flavor violating  couplings can also be read off Eq. (\ref{fulyuk}) as
\begin{eqnarray}
g_{h\tau\mu,h\mu\tau} = \frac{v^2}{\sqrt{2}\Lambda^2}\left(S_e^\dagger g_{ij} T_e \right)_{32,23}.
\label{gmt}
\end{eqnarray}
These couplings 
have been studied in connection with the CMS report \cite{Khachatryan:2015kon}
\begin{equation}
 B(h\to \mu\tau)=(0.84_{-0.37}^{+0.39})\% 
\end{equation}
with the resulting constraint $\sqrt{g_{h\tau\mu}^2+g_{h\mu\tau}^2}< 3.6\times 10^{-3}$.  LFV Higgs decays have been discussed using the effective Lagrangian framework by a number of authors \cite{Blankenburg:2012ex,Crivellin:2015mga,Harnik:2012pb,Belusca-Maito:2016axk,Dorsner:2015mja}. Comparing Eq. (\ref{gmt}) with Eq. (\ref{gtt}) and assuming the elements of $g_{ij}$ are of the same order of magnitude this implies that
\begin{equation}
\epsilon_\tau\sim\tilde{r}_\tau\lesssim 0.35
\label{cmscon}
\end{equation}
As illustrated by the green circle in Figure~\ref{fig}.
One can also assume that the matrix $g_{ij}$ is of the Fritzsch type \cite{Fritzsch:1979zq,Harnik:2012pb}, in which case $g_{23}/g_{33}\sim \sqrt{m_\mu/m_\tau}$ and the result would be less restrictive by a factor of four.

It is also possible to constrain these couplings from the measurement of the $h \to \tau\tau$ rate. From CMS  \cite{Khachatryan:2014jba} and ATLAS  \cite{Aad:2015vsa} we have
\begin{eqnarray}
\mu \equiv \frac{\sigma(h\to\tau\tau)}{\sigma(h\to\tau\tau)_{SM}}&=& 0.9\pm 0.28  {\rm ~CMS}\nonumber \\
&=& 1.43^{+0.43}_{-0.37} {\rm ~ATLAS}
\label{expc}
\end{eqnarray}
This can be compared with the rate calculated from the effective Lagrangian. With $\beta_\tau=\sqrt{1-4m_\tau^2/m_h^2}$ we find
\begin{eqnarray}
\Gamma= \frac{\beta_\tau}{8\pi m_H}m_\tau^2\left(\frac{m_H^2}{v^2}\right)\left(\beta_\tau^2|r_\tau|^2+|\tilde{r}_\tau|^2\right).
\end{eqnarray}
We plot the resulting constraints in Figure~\ref{fig} and compare them to the one from Eq. (\ref{cmscon}). The figure shows that these constraints allow the quantity $  (r_\tau \tilde{r}_\tau ) /( r_\tau^2 + \tilde{r}_\tau^2  )$ to take a maximum value of about 0.32.

It is well known that the simultaneous existence of scalar and pseudo-scalar couplings to a fermion pair as in Eq. (\ref{thisL}) signals CP violation.
The consequences of these couplings for the $\tau$ lepton  to Higgs have been analyzed some time ago \cite{He:1993fd}. 
In the standard treatment of this problem, one can define a density matrix $R$ for production of polarized tau-leptons
with polarization described by a unit polarization vector
${\bf n}_{\tau(\bar \tau)}$ in the $\tau(\bar \tau)$-rest frame. With the amplitude in
Eq. (\ref{thisL}) the CP violating part of the density matrix is given by
\begin{eqnarray}
   R_{CP}= -N  \beta_\tau {\rm Re}(r_\tau\tilde{r}_\tau) {\vec{p}_\tau}\cdot (
     {\bf n_\tau}\times {\bf n_{\bar \tau}})\;,
\end{eqnarray}
where $N$ is a normalization constant and 
${\vec{p}_\tau} $ is the three momentum
direction of the tau-lepton. Beyond tree-level, 
$r_\tau$ and $\tilde{r}_\tau$ acquire imaginary parts and the density matrix has additional terms that we will not consider in this paper. $R_{CP}$
contains all the information about experimental observables with the weak decay of the tau-leptons analyzing their polarization. 

The simplest mode to consider is the two body decay  
\begin{equation}
 \tau^- \rightarrow \pi^-\nu_{\tau},\ \
      \tau^+ \rightarrow \pi^+\bar \nu_{\tau}\;.
\end{equation}
Denoting $\vec{p}_{\pi^\pm}$ as the three- momenta of the pions 
 in the Higgs rest frame, a T-odd correlation sensitive to CP violation is given by
\begin{eqnarray}
{\cal O}_\pi = \vec{p}_\tau\cdot(\vec{p}_{\pi^+} \times \vec{ p}_{\pi^-}).
\end{eqnarray}
This can be measured by the integrated counting asymmetry
\begin{eqnarray}
A_\pi =\frac{N({\cal O}_\pi >0)-N({\cal O}_\pi <0) }{N({\cal O}_\pi >0)+N({\cal O}_\pi<0 )}=
     \frac{ \pi}{ 4} \beta_\tau 
  \frac{ (r_\tau \tilde{r}_\tau )}{\beta_\tau^2  r_\tau^2 + \tilde{r}_\tau^2  }.
   \label{piasym}
\end{eqnarray}

The three body leptonic decay $\tau^\pm \to \ell^\pm \nu\bar{\nu}$  can also be calculated analytically in  the limit $m_\tau<<m_H$, $\beta_\tau\to 1$ and $m_\ell << m_\tau$. The T-odd correlation 
\begin{eqnarray}
{\cal O}_\ell = \vec{p}_\tau\cdot(\vec{p}_{\ell^+} \times \vec{ p}_{\ell^-}),
\end{eqnarray}
where now $\vec{p}_{\ell^\pm}$ denotes the three-momenta of the charged lepton in the Higgs rest frame, can be measured with the integrated counting asymmetry 
\begin{eqnarray}
A_{\ell}&=& \frac{\pi}{36}\frac{r_\tau\ \tilde{r}_\tau}{|r_\tau|^2+|\tilde{r}_\tau|^2}.
\label{lepasym}
\end{eqnarray}

Although the calculation for the two modes discussed so far can be carried out analytically, it is convenient to  implement Eq. (\ref{thisL}) in  {\tt MADGRAPH5} \cite{Alwall:2014hca} with the aid of {\tt FEYNRULES} \cite{Christensen:2008py}. This allows us to verify numerically the results of Eqs. (\ref{piasym}) and (\ref{lepasym}). It will also allow us to discuss more complicated tau decay modes in a future publication. 

Combined with the limits from Figure~\ref{fig},  Eq. (\ref{piasym}) shows that a very large  asymmetry, of order $25\%$ is still allowed. Of course there are many complications that will reduce this asymmetry at LHC, such as it not being possible to reconstruct the Higgs rest frame in this mode, but a full phenomenological analysis is beyond the scope of this work. The asymmetries we discuss above may be measurable at a future linear collider. In any case, there are other methods to measure $\tilde{r}_\tau$ at the LHC and to estimate the potential sensitivity. Ref.~\cite{Berge:2015nua}, for example,  finds that $\tilde{r}_\tau$ can be determined with an uncertainty of .25 (.15) with 150 (500) fb$^{-1}$ . This implies that 150 fb$^{-1}$ would be enough to improve on the LFV constraint, Eq.~\ref{cmscon} within the scenario studied in this work. 

\begin{acknowledgments}

The work of G.V. was supported in part by the DOE under contract number DE-SC0009974. X-G He was supported in part by MOE Academic Excellent Program (Grant No.~102R891505) and MOST of ROC (Grant No.~MOST104-2112-M-002-015-MY3), and in part by NSFC (Grant Nos.~11175115 and 11575111) and Shanghai Science and Technology Commission (Grant No.~11DZ2260700) of PRC.  X.~G.~H. thanks Korea Institute for Advanced Study (KIAS) for their hospitality and partial support while this work was completed. We thank Brian Le for discussions on CP violation with taus in ATLAS.

\end{acknowledgments}

\end{document}